\documentclass[aps,prl,showpacs,twocolumn,superscriptaddress]{revtex4}

\begin{document}

\title{Reply to Comment on ``Analysis of the spatial distribution between successive
earthquakes''}

\author{J\"orn Davidsen}
\email[]{davidsen@phas.ucalgary.ca}\affiliation{Complexity Science
Group, Department of Physics \& Astronomy, University of Calgary,
Calgary, Alberta T2N 1N4, Canada }
\author{Maya Paczuski}
\email[]{maya.paczuski@ucalgary.ca} \affiliation{Complexity
Science Group, Department of Physics \& Astronomy, University of
Calgary, Calgary, Alberta T2N 1N4, Canada }

\date{\today}

\begin{abstract}
This is a reply to the Comment on ``Analysis of the spatial
distribution between successive earthquakes'' by Maximilian Jonas
Werner and Didier Sornette.
\end{abstract}


\maketitle

Werner and Sornette (WS)~\cite{werner07} claim to show that our
analysis of spatial distances between successive earthquakes in
southern California \cite{davidsen05m} does not allow to detect
length scales associated with aftershock zones. However, almost
all of their arguments are based on an analysis that is completely
insensitive to finite size scaling: A function $F$ obeys finite
size scaling if and only if a function $f$ and constants $\alpha$
and $\beta$ exist such that $F_L(x)=L^\alpha f(x/L^\beta)$ for any
(large) system size $L$. While we studied the probability density
function (PDF) $P_{L}(\Delta r)$ of spatial distances over all
boxes of a given linear size $L$ and its variation with $L$, WS do
not take into account any variation with box size. Instead, they
consider all events in the given catalog being in a single "box"
whose size and shape is determined by the catalog. Since the
scaling function $f$ is not constrained to a particular form
\emph{a priori}, the presence or absence of finite size scaling
cannot be established by considering a single scale $L$ as claimed
by WS. Thus, they cannot make any statements about the variation
with linear size $L$ and the existence of finite size scaling,
which is a crucial point in our line of argument regarding
aftershock zone scaling with main shock magnitude. In particular,
this is true for their results obtained for earthquake catalogs
from Northern California and Japan.

In fact, the results they obtained are strikingly similar to
results for periods of quasi-stationary seismic activity
\cite{corral06}. This suggests that the observation periods in
Japan and Northern California that WS analyzed show rates of
seismic activity which are rather homogeneous compared to the
period we analyzed for southern California (1984-2000). As we
emphasized in Ref.~\cite{davidsen05m}, the power law behavior of
$P_{L}(\Delta r)$, for earthquakes in Southern California, only
holds for \emph{very long} observation periods where the rate of
earthquake activity is highly heterogeneous in \emph{space and time}.
This was confirmed in a detailed study in Ref.~\cite{corral06}.
The particular combination of periods and regions with high
seismic activity and low seismic activity in southern California
is the reason for observing the scale-free PDF for spatial
distances. Thus, considering only the most active period around
the Landers event, or neglecting it, is expected to lead to strong
deviations in the PDF as observed by WS. In particular, their
results do not in any way contradict the results we found and our
discussion emphasizing the difference between homogeneous and
heterogeneous rates in~\cite{davidsen05m}.

Another point raised by WS is that aftershocks occur at distances 
larger than the main shock's rupture length. This is well-known,
yet the vast majority of what are typically considered aftershocks 
occur within distances which are comparable to and no more than 
a few times the rupture length (see, for example, 
Ref.~\cite{shcherbakov05d} and references therein). In particular,
the largest rupture length in the catalog from southern California 
we studied is $\approx 90$km. This is significantly less than the
spatial extent of the area studied ($\approx 500$km) which allows 
us to test the hypothesis of aftershock zone scaling with magnitude. 
As our results in Ref.~\cite{davidsen05m} show, no physical length 
scale exists in the range from $20$ km to $\approx 500$km.

WS also present results for a spatially extended version of the
ETAS model. While they claim that this extended model is an
accurate description of aftershock zone scaling with main shock
magnitude, a comparison of the Landers sequence shown in their
Fig.~1~\cite{werner07} and the ETAS model in Fig.~2~\cite{werner07}
suggests otherwise. More importantly, the particular form of 
$P(d)$ is speculative --- it does not follow from the work by Kagan
~\cite{kagan02} --- and it is by no means generally accepted. 
Despite some indication for a power law with $\mu \approx 1.35$ 
for \emph{short} times after the mainshock~\cite{felzer06}, 
other results even directly contradict the form of a power-law 
decay for distances larger than the main shock rupture length 
\cite{davidsen05pm,davidsen06pm} if the activity is considered 
over the long time scales relevant for our study in 
Ref.~\cite{davidsen05m}. To
summarize, the behavior of a model, which is not an accurate
description of aftershock zone scaling, cannot prove that our
earlier results are insensitive to the existence of physical
length scales associated with aftershock zones.

%
%

\end{document}